# Rotational modes of oscillation of rodlike dust grains in a plasma

M. P. Hertzberg, S. V. Vladimirov[*], and N. F. Cramer
*School of Physics, The University of Sydney, New South Wales 2006, Australia*

Three dimensional rotatory modes of oscillations in a one-dimensional chain of rodlike charged particles or dust grains in a plasma are investigated. The dispersion characteristics of the modes are analyzed. The stability of different equilibrium orientations of the rods, phase transitions between the different equilibria, and a critical dependence on the relative strength of the confining potential are analyzed. The relations of these processes with liquid crystals, nanotubing, and plasma coating are discussed.

PACS numbers: 52.35.Ra, 52.35.Kt, 52.25.Vy

## I. INTRODUCTION

Recently, there has been an increasing interest in the properties of structures involving colloidal charged particles levitating in a plasma [1–4]. The dynamic properties of the particle motion, formation of colloidal crystals and phase transitions in plasma-dust systems are important fundamental questions related to the general theory of self-organization in open dissipative systems [4]. The cases already studied, experimentally and theoretically, mostly correspond to spherical dust grains, but there is growing interest in the properties of colloidal structures composed of elongated (cylindrical) particles [5, 6] levitating in the sheath region of a gas discharge plasma. In these experiments, various arrangements of such grains, levitating horizontally (i.e., oriented parallel to the lower electrode and perpendicular to the gravity force) and vertically (i.e., oriented perpendicular to the lower electrode and parallel to the gravity force) have been observed.

It is necessary to stress that for elongated particles, properties of the plasma environment and especially the sheath properties are very important [6]. Indeed, the sheath region is characterized by strong inhomogeneities of the plasma parameters. In contrast to spherical grains which can practically be considered as point-like particles, rods can be affected by these inhomogeneities [6]. However, as a first approximation, we assume here the ambient plasma to be homogeneous; in the case of experimental levitation in the sheath this obviously restricts the possible rod length (it should be less than the characteristic inhomogeneity scale); however this approximation can have direct applications even for longer rods for the (possible future) experiments when the rod structures levitate in plasma bulk as, e.g., under micro-gravity conditions.

The oscillations of chains of point-like charged particles have previously been analyzed [7, 8]. Unlike point-like or spherical particles, elongated rotators exhibit a number of additional oscillations related to the new (rotational) degrees of freedom [9, 10]. Lattices composed of rodlike particles will therefore exhibit rotational oscillation modes, analogous to those existing in liquid crystals [11]. It is natural to expect that the excitation and interactions of all these modes will strongly affect the lattice dynamics, leading in particular to new types of phase transitions, as well as affecting those phase transitions already existing in lattices composed of spherical grains. Earlier we have briefly communicated the first results of the investigation of lattice oscillations in the one-dimensional chain consisting of rod-shaped particles [10] where characteristics of the modes associated with the motions and rotations of rods in the (vertical) plane of the chain were obtained. Here, we present a full three-dimensional analysis of the rotatory modes in the chain of rodlike particles, and analyze the critical dependence of the equilibrium and stability of such a chain on the external potential.

## II. BASIC EQUATIONS

Each rodlike particle is modelled as a rotator having two charges (and masses) concentrated on the ends of the rod, see Fig. 1, the upper charge being $Q_a$ and the lower charge $Q_b$. For further simplicity, we assume that the charges are constant and the masses are equal. The rod of length $L$, connecting these two charges, has zero radius and mass. We consider a one-dimensional infinite linear chain of rods, with their centers of mass evenly separated by the distance $d$ in equilibrium, along the $x$-axis. Note that the approximation of an *infinite* chain is reasonable for the several dozen or so dust particles that are involved in a typical experiment. The one-dimensional array (chain) of rods is also a good first approximation since, firstly, such chains can arise in specially designed experiments and it is therefore important to establish the role of a coupling potential in the horizontal $y$-direction on the stability of the chain, and, secondly, although the introduction of parallel chains would complicate the mode structure further, the main mode characteristics for perturbations propagating along the chain will survive at least in the linear approximation.

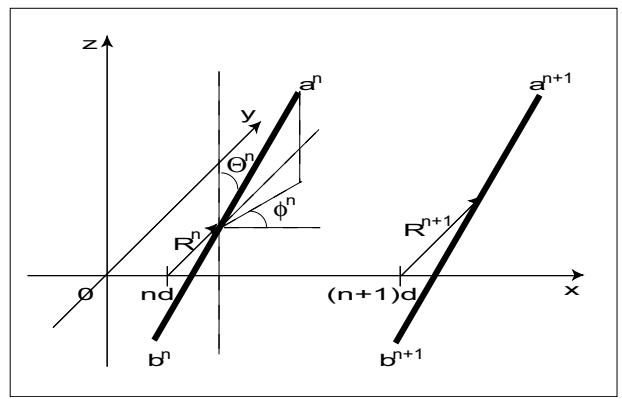

FIG. 1: Geometry of the rod.

[*]Email: S.Vladimirov@physics.usyd.edu.au; URL: http://www.physics.usyd.edu.au/~vladimi

The relevant forces are due to the external potentials and the interparticle interactions. The external potential $\Phi_{\text{ext}}$ is a combination of the potentials due to both gravitation and the external electrodes. The interparticle force is Coulombic in nature. However, since the dust grain is shielded by the surrounding plasma, there is an exponential decay of the interparticle potential with distance, as follows;

$$\Phi_a(\mathbf{r}) = \frac{Q_a}{4\pi\varepsilon_0|\mathbf{r}|} \exp\left(-\frac{|\mathbf{r}|}{\lambda_D}\right), \qquad (1)$$

where $\lambda_D$ is the plasma Debye length, which is the scale over which plasma shielding is effective and as such, is a function of the plasma parameters (temperature, density etc).

Hence, we can form the Lagrangian

$$\mathcal{L} = \frac{m}{2}\sum_n \left(\frac{d\mathbf{R}^n}{dt}\right)^2 + \frac{I}{2}\sum_n \left[\left(\frac{d\phi^n}{dt}\right)^2 (\sin\theta^n)^2 + \left(\frac{d\theta^n}{dt}\right)^2\right]$$
$$- \left(\sum_n Q_a\Phi_a + \sum_n Q_a\Phi_b + \sum_n Q_b\Phi_a + \sum_n Q_b\Phi_b + \sum_n Q_a\Phi_{\text{ext}} + \sum_n Q_b\Phi_{\text{ext}}\right), \qquad (2)$$

where the sum is over all the particles, and $I$ is the common rotational inertia of each particle. Here $\theta$ is the angle the rod makes with the $z$ (vertical) axis, and $\phi$ is the angle the projection of the rod onto the $x-y$ (horizontal) plane makes with the $x$-axis (the direction of the chain of particles), see Fig. 1. The first two terms on the right hand side correspond to the kinetic energy, and terms like $Q_a\Phi_a$ describe the interaction on the upper charge $Q_a$ of the $n^{\text{th}}$ particle with the potential $\Phi_a$ due to the upper charges $Q_a$ of all the other particles, etc. (The force between the upper and lower charges on a single particle is of course balanced by the stress in the rod connecting them). Note the summation implied in the calculation of $\Phi_a, b$ for the $n^{\text{th}}$ particle is, in principle, over all the other particles. However, we shall only allow the two nearest neighbor interactions in our approximation due to the screened Coulomb nature of the interparticle interaction. Indeed, we would not expect that highly charged rods can appear at distances significantly less than Debye length. Note that the separation between rods is an adjustable parameter in our calculations. This separation depends on the parameters of the plasma: in the experiments [6], it varies from 1 mm to 0.3 mm. This scale length is more than or of the order of the Debye length for the experiments, and this justifies our assumption of the nearest neighbor interactions.

Modes associated with the center of mass motion are analogous to those in chains of spherical particles [7, 8, 10] so here we investigate the rotational behavior, unique for rodlike particles, in detail. The Euler Lagrange equations of angular motion can then be written down for the $n^{\text{th}}$ particle,

$$\frac{d}{dt}\frac{\partial\mathcal{L}}{\partial\dot{\theta}^n} = \frac{\partial\mathcal{L}}{\partial\theta^n}, \qquad (3)$$

and similarly for $\phi^n$.

In order to compute these derivatives we must write down the displacement vectors from each charge to each neighboring charge. Let $\mathbf{R}^n$ locate the center of mass of the $n^{\text{th}}$ dust grain, relative to the equilibrium position. We then define $\mathbf{S}^n$ to be the direction of the upper charge relative to the center of mass, so in spherical co-ordinates

$$\mathbf{S}^n = (\cos\phi^n \sin\theta^n, \ \sin\phi^n \sin\theta^n, \ \cos\theta^n). \qquad (4)$$

Hence, the $n^{\text{th}}$ upper charge $Q_a$ is at

$$\mathbf{a}^n = nd\mathbf{e}_x + \mathbf{R}^n + \frac{L}{2}\mathbf{S}^n, \qquad (5)$$

and the lower charge $Q_b$ is at

$$\mathbf{b}^n = nd\mathbf{e}_x + \mathbf{R}^n - \frac{L}{2}\mathbf{S}^n. \qquad (6)$$

We may now define the four displacement vectors *from* the charges on the $(n+1)^{\text{th}}$ grain *to* the charges on the

$n^{\text{th}}$ grain as follows,

$$\mathbf{r}_{aa}^{n+} = -d\mathbf{e}_x + (\mathbf{R}^n - \mathbf{R}^{n+1}) + \frac{L}{2}(\mathbf{S}^n - \mathbf{S}^{n+1}), \quad (7)$$

$$\mathbf{r}_{ba}^{n+} = -d\mathbf{e}_x + (\mathbf{R}^n - \mathbf{R}^{n+1}) + \frac{L}{2}(\mathbf{S}^n + \mathbf{S}^{n+1}), \quad (8)$$

$$\mathbf{r}_{bb}^{n+} = -d\mathbf{e}_x + (\mathbf{R}^n - \mathbf{R}^{n+1}) - \frac{L}{2}(\mathbf{S}^n - \mathbf{S}^{n+1}), \quad (9)$$

$$\mathbf{r}_{ab}^{n+} = -d\mathbf{e}_x + (\mathbf{R}^n - \mathbf{R}^{n+1}) - \frac{L}{2}(\mathbf{S}^n + \mathbf{S}^{n+1}). \quad (10)$$

Similar equations exist for the vectors from the charges on the $(n-1)^{\text{th}}$ grain to the charges on the $n^{\text{th}}$ grain.

The external potential can be approximated by a parabolic potential for small oscillations, whose minimum lies at the center of mass of a rod ($y=0$, $z=0$). The assumption of an infinite chain in the $x$-direction removes the need for a confining potential in that direction. The external potential is therefore assumed to act in both the $z$ (vertical) and $y$ directions, such that the potential energy of the rod is, with coordinates , is

$$Q_a \Phi_{\text{ext}}(\mathbf{a}^n) + Q_b \Phi_{\text{ext}}(\mathbf{b}^n) = k_y y^2 + k_z z^2, \quad (11)$$

where $(y,z)$ are the coordinates of the upper charge $Q_a$. Note that the $y$-component of the external potential is purely electrical and the $z$-component is a combination of the electric and gravitational potentials.

Use of the chain rule now gives rise to the following equation of motion for $\theta^n$, including only the two nearest neighbor particle interactions:

$$\begin{aligned}
I\left[\ddot{\theta}^n - \left(\dot{\phi}^n\right)^2 \sin\theta^n \cos\theta^n\right] = -\frac{Q_a L}{2} &\Bigg[ \frac{\Phi'_a(\mathbf{r}_{aa}^{n+})}{|\mathbf{r}_{aa}^{n+}|} \left(r_{aa,x}^{n+}\cos\phi^n\cos\theta^n + r_{aa,y}^{n+}\sin\phi^n\cos\theta^n - r_{aa,z}^{n+}\sin\theta^n\right) \\
&+ \frac{\Phi'_a(\mathbf{r}_{aa}^{n-})}{|\mathbf{r}_{aa}^{n-}|} \left(r_{aa,x}^{n-}\cos\phi^n\cos\theta^n + r_{aa,y}^{n-}\sin\phi^n\cos\theta^n - r_{aa,z}^{n-}\sin\theta^n\right) \\
&+ \frac{\Phi'_b(\mathbf{r}_{ba}^{n+})}{|\mathbf{r}_{ba}^{n+}|} \left(r_{ba,x}^{n+}\cos\phi^n\cos\theta^n + r_{ba,y}^{n+}\sin\phi^n\cos\theta^n - r_{ba,z}^{n+}\sin\theta^n\right) \\
&+ \frac{\Phi'_b(\mathbf{r}_{ba}^{n-})}{|\mathbf{r}_{ba}^{n-}|} \left(r_{ba,x}^{n-}\cos\phi^n\cos\theta^n + r_{ba,y}^{n-}\sin\phi^n\cos\theta^n - r_{ba,z}^{n-}\sin\theta^n\right) \Bigg] \\
&+ \frac{Q_b L}{2}[a \leftrightarrow b] - Q_a \frac{\partial \Phi_{\text{ext}}(\mathbf{a}^n)}{\partial \theta^n} - Q_b \frac{\partial \Phi_{\text{ext}}(\mathbf{b}^n)}{\partial \theta^n}, \quad (12)
\end{aligned}$$

where $\Phi'(\mathbf{r}) = d\Phi/d|\mathbf{r}|$, and where $[a \leftrightarrow b]$ is shorthand for another set of terms which are identical to the first set in square brackets on the right hand side but with $a$ replaced by $b$.

The corresponding equation for the azimuthal angle $\phi^n$ is

$$\begin{aligned}
I\sin\theta^n\left(\ddot{\phi}^n\sin\theta^n + 2\dot{\phi}^n\cos\theta^n\dot{\theta}^n\right) = -\frac{Q_a L}{2} &\Bigg[ \frac{\Phi'_a(\mathbf{r}_{aa}^{n+})}{|\mathbf{r}_{aa}^{n+}|} \left(-r_{aa,x}^{n+}\sin\phi^n\sin\theta^n + r_{aa,y}^{n+}\cos\phi^n\sin\theta^n\right) \\
&+ \frac{\Phi'_a(\mathbf{r}_{aa}^{n-})}{|\mathbf{r}_{aa}^{n-}|} \left(-r_{aa,x}^{n-}\sin\phi^n\sin\theta^n + r_{aa,y}^{n-}\cos\phi^n\sin\theta^n\right) \\
&+ \frac{\Phi'_b(\mathbf{r}_{ba}^{n+})}{|\mathbf{r}_{ba}^{n+}|} \left(-r_{ba,x}^{n+}\sin\phi^n\sin\theta^n + r_{ba,y}^{n+}\cos\phi^n\sin\theta^n\right) \\
&+ \frac{\Phi'_b(\mathbf{r}_{ba}^{n-})}{|\mathbf{r}_{ba}^{n-}|} \left(-r_{ba,x}^{n-}\sin\phi^n\sin\theta^n + r_{ba,y}^{n-}\cos\phi^n\sin\theta^n\right) \Bigg] \\
&+ \frac{Q_b L}{2}[a \leftrightarrow b] - Q_a \frac{\partial \Phi_{\text{ext}}(\mathbf{a}^n)}{\partial \phi^n} - Q_b \frac{\partial \Phi_{\text{ext}}(\mathbf{b}^n)}{\partial \phi^n}. \quad (13)
\end{aligned}$$

Note that these equations are extremely nonlinear, involving sine and cosine dependence on angles, and exponential and inverse distance dependence contained implicitly in the potential terms. Moreover, the $\theta$ and $\phi$ behavior is coupled. In order to proceed we must *linearize* these equations about an equilibrium position. We shall call the equilibrium colatitudinal and azimuthal angles, about which we consider small perturbations, $\theta_0$ and $\phi_0$ respectively, which are common to all particles. The analysis then proceeds as follows: let $\epsilon$ be a small perturbation of $\theta$, so that $\theta^n = \theta_0 + \epsilon^n$; we may then approximate

$$\cos\theta^n = \cos\theta_0 - \sin\theta_0 \epsilon^n, \quad (14)$$

$$\sin\theta^n = \sin\theta_0 + \cos\theta_0 \epsilon^n, \quad (15)$$

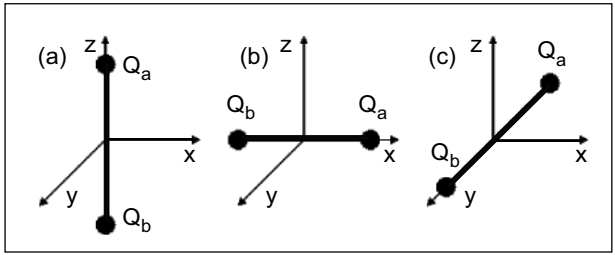

FIG. 2: The three equilibrium orientations of the rodlike dust grains.

and similarly for $\phi$ by letting $\eta$ be a small perturbation from $\phi_0$. Note that the linearising process for terms like $\Phi_b'(\mathbf{r}_{ba}^{n+})/|\mathbf{r}_{ba}^{n+}|$ in fact involves taking the zeroth and first order terms in a Taylor expansion of six variables, namely $\epsilon^n$, $\epsilon^{n+1}$, $\epsilon^{n-1}$, $\eta^n$, $\eta^{n+1}$, $\eta^{n-1}$. The full result for $\theta^n$ is long and complicated, and is written down in the Appendix, Eqn. (31). We simply note that it is coupled and has only a numerical solution.

We next determine the common *equilibrium* orientation of the particles. Thus we set $\epsilon^n = \epsilon^{n+1} = \epsilon^{n-1} = \eta^n = \eta^{n+1} = \eta^{n-1} = 0$, and let the acceleration $\ddot{\theta}^n$ be zero. We find

$$\frac{Q_a L}{2}\left[\frac{\Phi_b'(\mathbf{r}_{ba}^{n+})}{|\mathbf{r}_{ba}^{n+}|} - \frac{\Phi_b'(\mathbf{r}_{ba}^{n-})}{|\mathbf{r}_{ba}^{n-}|}\right]d\cos\phi_0\cos\theta_0 \ - \ [a\leftrightarrow b]$$
$$-Q_a\frac{\partial\Phi_{\text{ext}}(\mathbf{a}^n)}{\partial\theta^n} - Q_b\frac{\partial\Phi_{\text{ext}}(\mathbf{b}^n)}{\partial\theta^n} \ = \ 0, \quad (16)$$

where each term is to be evaluated at the relevant equilibrium orientation. This tells us the form of external potential required for an equilibrium orientation of $(\theta_0,\phi_0)$. However, as indicated earlier, the external potential of relevance to us is parabolic about the center of mass, equation (11).

Thus the external potential derivatives at each charge should vanish at the equilibrium; this occurs only at the $y$ and $z$ axes for a potential of the form (11), unless $k_y = k_z$. Similarly we must ensure

$$\left[\frac{\Phi_b'(\mathbf{r}_{ba}^{n+})}{|\mathbf{r}_{ba}^{n+}|} - \frac{\Phi_b'(\mathbf{r}_{ba}^{n-})}{|\mathbf{r}_{ba}^{n-}|}\right]d\cos\phi_0\cos\theta_0 = 0. \quad (17)$$

The result is that equilibria exist in only the following orientations (unless $k_y = k_z$); $(\theta_0,\phi_0) = (0,\phi_0)$ (with $\phi_0$ arbitrary) or $(\pi/2,0)$ or $(\pi/2,\pi/2)$, which we draw successively as shown in Fig. 2 (a), (b), and (c), respectively. If $k_y = k_z$, equilibrium exists for the orientation $(\theta_0,\pi/2)$, where $\theta_0$ is arbitrary.

Recall that the chain of rods is located along the $x$-axis, so that the first and the third equilibria are essentially the same, the only difference lying in the interaction with the external potential. The analysis of oscillations about these equilibria shall occupy the bulk of this paper. Oscillations in $\theta$ about the second equilibrium have already been considered in Ref. [10], and oscillations in $\phi$ about that equilibrium can be obtained simply by exchanging $k_y$ and $k_z$. Since $\phi$ is undefined for a vertically oriented rod, we concentrate here on the horizontal equilibrium case $(\pi/2,\pi/2)$.

### III. AZIMUTHAL MOTION

The perturbation equation describing small oscillations in azimuthal angle $\phi^n$ about the equilibrium Fig. 2(c), $\theta = \pi/2$, $\phi = \pi/2$, can be obtained from Eqn. (31) (see Appendix) for the $\theta$ perturbation equation by rotating the $y$- and $z$-axes about the $x$-axis, so the new $z$-axis lies along the old $y$-axis. The equilibrium is now, in terms of the new polar angles, $(\theta_0' = 0, \phi_0' = 0)$, and the perturbation $\eta$ in the old angle $\phi$ is the negative of the perturbation $\eta'$ in the new angle $\theta'$. The resulting equation for $\eta$ is

$$I\ddot{\eta}^n =$$
$$-\frac{L^2}{4}\left[Q_a\Phi_a''(d) - \frac{Q_a}{L_d}\Phi_b'(L_d)\right](2\eta^n - \eta^{n+1} - \eta^{n-1}) -$$
$$\frac{d^2L^2}{4}\left[\frac{Q_a}{L_d^2}\Phi_b''(L_d) - \frac{Q_a}{L_d^3}\Phi_b'(L_d)\right](2\eta^n + \eta^{n+1} + \eta^{n-1})$$
$$+ [Q_a \leftrightarrow Q_b] + \frac{L^2}{2}k_y\eta^n, \quad (18)$$

where $L_d^2 = L^2 + d^2$. Thus the $\epsilon$ and $\eta$ behavior decouples. (In fact they decouple in all the above three equilibrium cases.) Note the presence of $2\eta^n - \eta^{n+1} - \eta^{n-1}$ in the first term. This term will vanish if all the rods rotate together. If we consider the upper charges $Q_a$ all moving together, it is apparent that this interaction should vanish, as a result of the $(n-1)^{\text{th}}$ and $(n+1)^{\text{th}}$ charges pushing in equal and opposite directions. However in the second term, the term $2\eta^n + \eta^{n+1} + \eta^{n-1}$ will not vanish. This is a consequence of the fact that as the rods rotate, the cross interaction between $Q_a$ and $Q_b$ (on adjacent rods) increases as they come closer together. The $(n+1)^{\text{th}}$ and $(n-1)^{\text{th}}$ contributions become unbalanced.

To investigate the existence of an oscillatory solution we compute the Fourier transform of Eq. (18), which gives the frequency $\omega$ as a function of the wavenumber $k$ in the $x$-direction:

$$I\omega^2 = + L^2\left[Q_a\Phi_a''(d) - \frac{Q_a}{L_d}\Phi_b'(L_d)\right]\sin^2(kd/2)$$
$$+ \frac{d^2L^2}{L_d^2}\left[Q_a\Phi_b''(L_d) - \frac{Q_a}{L_d}\Phi_b'(L_d)\right]\cos^2(kd/2)$$
$$+ [Q_a \leftrightarrow Q_b] - \frac{L^2}{2}k_y. \quad (19)$$

Now the Debye-Coulomb potential in equation (19) has derivatives,

$$\Phi_a'(\mathbf{r}) = -\frac{Q_a}{4\pi\varepsilon_0|\mathbf{r}|^2}\left(1 + \frac{|\mathbf{r}|}{\lambda_D}\right)\exp\left(-\frac{|\mathbf{r}|}{\lambda_D}\right), \quad (20)$$

$$\Phi_a''(\mathbf{r}) = \frac{Q_a}{4\pi\varepsilon_0|\mathbf{r}|^3}\left(2 + 2\frac{|\mathbf{r}|}{\lambda_D} + \frac{|\mathbf{r}|^2}{\lambda_D^2}\right)\exp\left(-\frac{|\mathbf{r}|}{\lambda_D}\right). \quad (21)$$

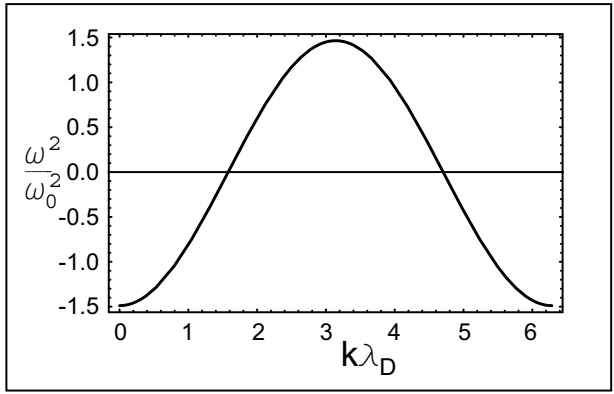

FIG. 3: Normalized frequency squared versus normalized wavenumber for perturbations in the angle $\phi$, travelling in the $x$ direction. The equilibrium orientation is $(\theta_0 = \pi/2, \phi_0 = \pi/2)$. Here $\omega_0$ is the dust plasma frequency, $d = \lambda_D$ and $k_y = 10^{-10} \text{kgs}^{-2}$.

Hence the product $Q_a\Phi_a'$ is negative, while $Q_a\Phi_a''$ is positive, as is true for any potential that falls off with distance. The result is that the coefficients of the oscillatory sine and cosine terms in Eq. (19) are always positive. Thus the dust particles would always exhibit stable oscillations, except for the presence of the term $-k_y L^2/2$ from the external potential, which acts to pull the grains away from this equilibrium to the $x$-axis (where $y = 0$). These competing terms may then give rise to regions of stable behavior and regions of unstable behavior. Recall that the moment of inertia is $I = mL^2/2$. Hence the factor $L^2$ cancels throughout and is only present, implicitly, through the quantity $L_d$. We may plot the dispersion relation, by selecting some typical values of the parameters involved: $m = 10^{-15}$kg, $Q = 10^3 e$ to $10^4 e$, and $\lambda_D = 300\mu$m. For the case $d \approx \lambda_D$, and for the particular choice of the external potential parameter $k_y = 10^{-10}$kg s$^{-2}$ (which can be controlled in an experiment), the resulting dispersion relation is shown in Fig. 3.

Note that the vertical axis is for the square of the normalized frequency $(\omega/\omega_0)^2$, where $\omega_0$ is a typical dust plasma frequency with $\omega_0^2 = 3Q^2/(4\pi\varepsilon_0 m\lambda_D^3)$, which is normally a few hundred radians per second. The horizontal axis is a normalized wavenumber $k\lambda_D$ for the first Brillouin zone. This plot clearly shows stable regions, corresponding to $\omega^2 > 0$, and unstable regions corresponding to $\omega^2 < 0$ (i.e $\omega$ imaginary.)

For a typical choice of the wavenumber, we can plot $(\omega/\omega_0)^2$ as a function of a normalized interparticle distance $d/\lambda_D$, (since $d$ is typically of the order of the Debye length), as shown in Fig. 4. Clearly at close interparticle distances the motion is stable, as the particles oscillate under their mutual repulsion. However at large distances the external potential dominates, and the motion becomes unstable. This is a result of the interparticle force decreasing with distance. Note that this instability will *always* arise at large distances, if $k_y > 0$. We reiterate that this horizontal motion is identical to the vertical case under the interchange $k_y \leftrightarrow k_z$.

It is useful to consider the case of small sized rods compared with the interparticle separation, $L \ll d$. Then

$$L_d = d\sqrt{1 + \left(\frac{L}{d}\right)^2} \approx d\left[1 + \frac{1}{2}\left(\frac{L}{d}\right)^2\right]. \quad (22)$$

It then makes sense to take $Q_a = Q_b$ since $Q = Q(z)$ for grains in a vertical sheath potential, and the upper and lower charges will essentially be at the same vertical position. One finds

$$\begin{aligned}\frac{1}{4}m\omega^2 =\ & Q_a\left[\Phi_a''(d) - \frac{1}{d}\Phi_a'(d) - \frac{1}{2}\frac{L^2}{d^2}\Phi_a''(d)\right.\\ & \left.+ \frac{1}{2}\frac{L^2}{d^3}\Phi_a'(d)\right] + Q_a\frac{L^2}{d}\left[\frac{1}{d^2}\Phi_a'(d) - \frac{1}{d}\Phi_a''(d)\right.\\ & \left.+ \frac{1}{2}\Phi_a'''(d)\right]\cos^2(kd/2) - \frac{1}{4}k_y. \quad (23)\end{aligned}$$

Here the first term on the right hand side is independent of wavenumber, whilst the second is not. We can understand this physically as follows: When $(L/d)$ is small, a rotation of the $n^{\text{th}}$ dust grain sets up oscillations since it becomes closer to the center of mass (and hence the dust particle on average) of the relevant nearest neighbor, regardless of the neighbor's orientation. This is embodied in the first term. Of secondary importance is the actual orientation of the neighbor, which is seen in the term involving wavenumber: note the $L^2$ factor, which shows that this correction is small.

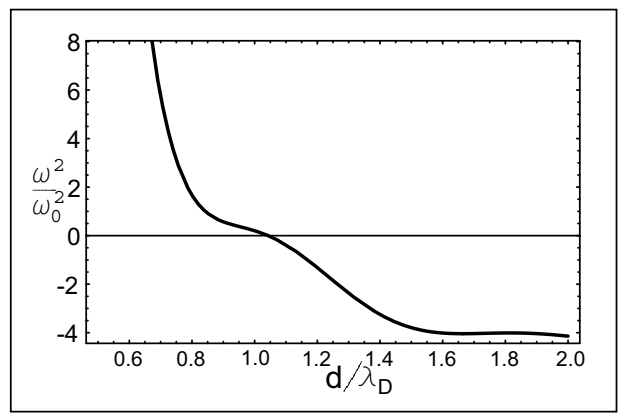

FIG. 4: Normalized frequency squared versus interparticle distance for perturbations in the angle $\phi$, travelling in the $x$ direction. Here $k\lambda_D = 8$, and the other parameters are as for Fig. 3.

## IV. COLATITUDINAL MOTION

We now investigate the behavior of the perturbation $\epsilon$ of the colatitudinal angle $\theta$, recalling that $\epsilon$ and $\eta$ completely decouple in the linear approximation. The equation of motion in the second (horizontal) equilibrium case is,

$$I\ddot{\epsilon}^n = -\frac{L^2}{4}\frac{Q_a}{d}\Phi'_a(d)(2\epsilon^n - \epsilon^{n+1} - \epsilon^{n-1})$$
$$+ \frac{L^2}{4}\frac{Q_a}{L_d}\Phi'_b(L_d)(2\epsilon^n - \epsilon^{n+1} - \epsilon^{n-1})$$
$$+ [Q_a \leftrightarrow Q_b] - \frac{L^2}{2}(k_z - k_y)\epsilon^n. \quad (24)$$

Note that this time *all* the dust particle interaction terms are of the form $2\epsilon^n - \epsilon^{n+1} - \epsilon^{n-1}$ and so the net force (apart from external influences) is zero if the rods rotate together. This expresses the fact that each plane (given by $\phi_0 \equiv \pi/2$, and all $\theta_0$ equal) is identical. Moreover, in the absence of an effective external potential (one in which $k_y = k_z$) there would exist equilibria at any $\theta_0$ value, since we would have no preferred direction. The stability of cases such as these will be pursued in the next section.

Fourier transforming Eq. (24) leads to the following dispersion relation:

$$I\omega^2 = L^2\left[\frac{Q_a}{d}\Phi'_a(d) - \frac{Q_a}{L_d}\Phi'_b(L_d)\right]\sin^2(kd/2)$$
$$+ [Q_a \leftrightarrow Q_b] + \frac{L^2}{2}(k_z - k_y). \quad (25)$$

Once again note the oscillatory dependence on wavenumber. The first term on the right hand side is this time *negative*, since $Q\Phi'(\mathbf{r}) < 0$ and $d < L_d$. Thus the dust particle's mutual repulsion causes instability. This results from the $Q_a\Phi_a$ interaction pushing away from equilibrium, dominating the cross interaction $Q_a\Phi_b$ that pushes back. Once again it is the fact that the dust grain's Debye-Coulomb potential falls off with distance that a net force results. Note the competing terms from the external potential. In the horizontal case one needs $k_z > k_y$ for the possibility of stability (of course wavenumber gaps are still possible). Note that similar dependence exhibit pairs of unbound spherical particles levitating in the confining potential in $x$- and $z$-directions [12]. We also recall that the interchange $k_z \leftrightarrow k_y$ gives the vertical equilibrium case. By selecting $k_z > k_y$ the dispersion relation will be qualitatively the same as in Fig. 3.

However, the behavior of the frequency as a function of interparticle distance shows a noticeable difference to the azimuthal oscillation case, as shown in Fig. 5. Note the reversed characteristic, where stability now increases as the particles get further apart. We explain this as follows: at close distances the motion is highly unstable, since the repulsion due to the $Q_a\Phi_a$ and $Q_b\Phi_b$ terms is so strong. However in the far region, it is the sign of $k_z - k_y$ that determines stability. Again by taking $L \ll d$ we approximate;

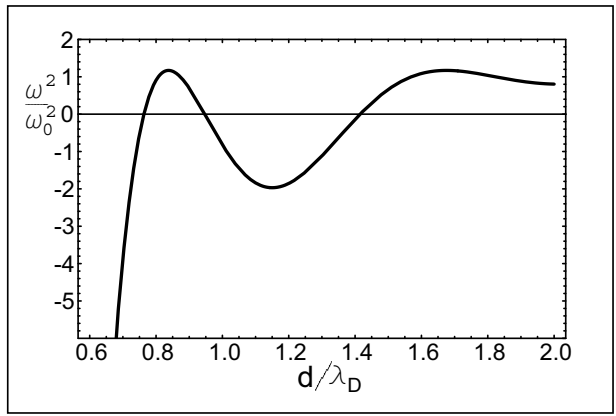

FIG. 5: Normalized frequency squared versus interparticle distance for perturbations in the angle $\theta$, travelling in the $x$ direction. Here $k\lambda_D = 8$.

$$\frac{1}{2}m\omega^2 = -Q_a\frac{L^2}{d^2}\left[\Phi''_a(d) - \frac{1}{d}\Phi'_a(d)\right]\sin^2(kd/2)$$
$$+ \frac{1}{2}(k_z - k_y). \quad (26)$$

We see that the oscillations become solely due to the confining potential as $L \to d$. In other words, the two competing terms in the coefficient of $\sin^2(kd/2)$ in equation (26) have cancelled one another. Note however that it is not strictly valid to let $L = 0$, since we cannot meaningfully discuss rotation of a rod with no length, and we have already divided by $L^2$.

## V. INSTABILITY

We now address the important issue of what happens if the motion is unstable. Instability implies that the rods may switch from the horizontal to the vertical equilibrium (or vice versa) depending on the sign of $k_z - k_y$. However, our analysis is only valid for small perturbation angles, and breaks down at large amplitudes. In order to examine what ensues if the rods lie near an unstable orientation we must consider what will happen physically: the rods will in fact move in opposite directions to move away from one another. Thus on the average, the even rods, say, will move clockwise and the odd rods counterclockwise (or vice versa). Hence it may be that there exists some intermediate value of stability between the vertical and horizontal. Our analysis is different now from earlier, since we are allowing alternate rods to have opposite equilibrium orientations. By considering the vertical case, this corresponds to equilibria at $\theta_0$ and $-\theta_0$ alternately. Let the even rods be described by perturbations $\epsilon^n$ in $\theta$ as before, and the odd neighbors by $\zeta^{n-1}$ and $\zeta^n$.

The equilibrium condition is now found to be given by

$$\sin(2\theta_0)\left[ -\frac{Q_a\Phi'_a(R_1) + Q_b\Phi'_b(R_1)}{2R_1} + \frac{Q_b\Phi'_b(R_2)}{R_2} - \frac{k_y - k_z}{8}\right] = 0, \quad (27)$$

where $R_1 = \sqrt{d^2 + L^2\sin^2\theta_0}$, and $R_2 = \sqrt{d^2 + L^2\cos^2\theta_0}$. The first order perturbation equation is

$$\begin{aligned}I\ddot{\zeta}^n &= \frac{L^2}{4}\left[-\frac{Q_a\Phi'_a(R_1) + Q_b\Phi'_b(R_1)}{R_1} + \frac{2Q_a\Phi'_b(R_2)}{R_2}\right] \\ &\times \left(2\cos(2\theta_0)\zeta^n - \eta^n - \eta^{n-1}\right) \\ &- \frac{L^2}{4}(k_y - k_z)\cos(2\theta_0)\zeta^n.\end{aligned} \quad (28)$$

Thus equilibrium occurs if $\theta_0 = 0$ (the vertical case), or $\theta_0 = \pi/2$ (the horizontal case), or if the term in brackets in (27) vanishes. In the case where $k_z = k_y$, and when $Q_a = Q_b$, an equilibrium occurs at $\theta_0 = \pi/4$, when the charges are furthest away from one another, and neighbors are out of phase by $\pi/2$. In this case the right hand side of the perturbation equation (28) vanishes; this is because for an external potential symmetric about the $x$ axis, an arbitrary rotation of $\theta_0$ may be made, as long as neighbors are at right angles to each other. If the potential is not symmetric ($k_y \neq k_z$), equilibria may occur at intermediate angles different to $\theta_0 = \pi/4$.

Hence the described system of rods can undergo phase transitions from one state to another i.e vertical to horizontal and vice versa, or to an intermediate equilibrium, dependent on the easily adjustable external potentials. In fact it is a natural extension to see that this is also true of the third equilibrium case mentioned in Section II, corresponding to the equilibrium at $(\pi/2, 0)$. Transitions between the vertical and horizontal equilibria have been observed in experiments [6]. This process has a relation to processes in liquid crystals [11], where a state of matter exists between the solid and liquid phases, wherein rod shaped molecules exhibit a partial alignment, rather than a rigid array seen in crystals. The direction of this partial alignment (and phase) can be altered by an external influence.

## VI. DISCUSSION AND CONCLUSIONS

The above analysis has ignored an important real feature of plasmas: friction. The motion of dust particles through the plasma medium will induce a resistive response from the plasma. Friction's importance depends on the plasma parameters. If the ionization percentage is low then the friction due to neutrals will dominate. If the ionization is high, then the friction from ions becomes important also. Now since the Lagrangian formulation cannot deal with non-conservative forces such as friction, we can only include friction after we have ascertained the equations of motion.

Thus, in the linear approximation, we may include a generic $\gamma\dot{\theta}$ term, where $\gamma$ is the frictional constant. Thus equation (24) becomes

$$\begin{aligned}I\ddot{\epsilon}^n + \gamma\dot{\epsilon}^n =& -\frac{L^2}{4}\frac{Q_a}{d}\Phi'_a(d)(2\epsilon^n - \epsilon^{n+1} - \epsilon^{n-1}) \\ &+ \frac{L^2}{4}\frac{Q_a}{L_d}\Phi'_b(L_d)(2\epsilon^n - \epsilon^{n+1} - \epsilon^{n-1}) \\ &+ [Q_a \leftrightarrow Q_b] - \frac{L^2}{2}(k_z - k_y)\epsilon^n.\end{aligned} \quad (29)$$

and the new frequency $\bar{\omega}$ expressed in terms of the old one becomes

$$\bar{\omega} = -i\frac{\gamma}{2I} + \sqrt{\omega^2 - \gamma^2/4I^2} \quad (30)$$

where the term $-i\gamma/2I$ is associated with damping, and $\sqrt{\omega^2 - \gamma^2/4I^2}$ shows a decrease in frequency. Note critical damping occurs when $\omega^2 = \gamma^2/4I^2$. Further generalizations of the main results taking into account friction are straightforward. We note that the actual friction with neutral particles is geometry dependent. In the case of the adopted model of a rotator with two spherical ball-like particles connected by the infinitely thin rod the friction constant *gamma* which is determined by the neutral gas pressure and the cross-section (i.e. geometry) is just twice that of the single spherical particle. In the case of the real rod geometry with finite rod radius the friction constant will naturally depend on the particular geometry. Note that the wave damping is also determined by the moment of inertia as can be seen from Eq. (30).

We began by noting that in a plasma, dust particles usually acquire a negative charge, and so can levitate if a negatively biased electrode is placed below them. The case of a one-dimensional chain of levitating rod shaped particles was investigated; we found that the different rotational modes of oscillation decoupled in the linear approximation. The specific behavior of the modes was analyzed through the dispersion relations. An oscillatory dependence on wavenumber was found, and a critical dependence on the relative strengths of the confining potential. The characteristic frequency range for these oscillatory modes is of the order of the dust plasma frequency $\omega_0$, as can be seen from Figs. 3–5. We note that the oscillations of rodlike particles also show modes associated with the motion of the center of mass of the rods. Dispersion characteristics of these modes are similar to those of spherical particles of comparable mass and charge, see [10], namely, the optic-mode like "bending" mode related to vertical motions of the centers of mass, and the acoustic mode related to horizontal (in the direction of the mode propagation) motions of the centers of mass. The characteristic frequency of the first mode is determined by the confining potential in the vertical direction; in experiments, and this frequency can be of the order of or higher than the dust plasma frequency depending on the particular experimental conditions [13]. The characteristic frequency range of the acoustic dust-lattice mode is lower, ranging from zero to $\omega_0$. The azimuthal and co-latitudinal modes showed opposite characteristics in the near and far interparticle distance regimes, respectively. The rods were then shown to move, or switch to the relevant equilibrium, dependent on the confining parameters. This is an example of a phase change phenomenon which is analogous to that observed in liquid crystals. The inclusion of the resistive effect of friction was an immediate and straightforward extension.

The ability to line up rods in different directions, by alternating the relative sizes of the confining potentials, is a powerful tool. This can be of use in, for example, plasma coating, if rod shaped objects are used as the basis to give strength to a material. The nanotube industry is another new area where this may find application. The elongated shape of these carbon based molecules shows an obvious connection to our discussion of 'rods', although the particular rod shape analyzed in this paper is just a first approximation to the true geometry of a nanotube [14].

We have neglected here some other plasma effects, such as the production of wakes due to the ion flow to the electrode. Since we consider dust charge interactions primarily in the horizontal plane, the modification of the Debye potential by the wake is only weak. We note that the ion wake effects can cause instabilities in the horizontal lattice wave propagation; for spherical particles this was shown in Ref. [15]. It is natural to expect a similar type of instability for the arrays of rods, and this is an interesting topic for future research. Also, for vertically oriented chains of dust charges, the wake potential has large effects (for chains of spherical grains, this was demonstrated in Ref. [16]). Finally, for structures levitating in the plasma bulk (as e.g. under micro-gravity conditions), the ion flow is absent.

There are several possibilities for further work. Firstly, the charge is in fact a function of vertical height, $Q = Q(z)$. Next, the extension to a 2 or 3 dimensional array may lead to further interesting results. An important extension is to a general charge distribution along the rod, rather than the two-point particles we have considered. Finally, by including higher order terms in the analysis, nonlinear waves and excitations as well as their interactions can be investigated in such chains.

## ACKNOWLEDGMENT

This work was supported by the Australian Research Council.

## APPENDIX

The 1st order equation of motion for $\epsilon$, about an arbitrary position $(\theta_0, \phi_0)$ is as follows:

$$\begin{aligned}
I\ddot{\epsilon}^n = & -\frac{Q_a L}{2}\Bigg[\left(\frac{L}{2}\Phi_a''(d) - \frac{L}{2d}\Phi_a'(d)\right)\left(\cos\phi_0\cos\theta_0(2\epsilon^n - \epsilon^{n+1} - \epsilon^{n-1})\right. \\
& \left. - \sin\phi_0\sin\theta_0(2\eta^n - \eta^{n+1} - \eta^{n-1})\right)\cos\phi_0\cos\theta_0 + \frac{L}{2d}\Phi_a'(d)(2\epsilon^n - \epsilon^{n+1} - \epsilon^{n-1}) \\
& + \frac{\Phi_b'(\mathbf{r}_{ba}^{n+})}{|\mathbf{r}_{ba}^{n+}|}\left(-d\cos\phi_0\cos\theta_0 + d\cos\phi_0\sin\theta_0\epsilon^n + d\sin\phi_0\cos\theta_0\eta^n - \frac{L}{2}(\epsilon^n - \epsilon^{n+1})\right) \\
& + \frac{\Phi_b'(\mathbf{r}_{ba}^{n-})}{|\mathbf{r}_{ba}^{n-}|}\left(d\cos\phi_0\cos\theta_0 - d\cos\phi_0\sin\theta_0\epsilon^n - d\sin\phi_0\cos\theta_0\eta^n - \frac{L}{2}(\epsilon^n - \epsilon^{n-1})\right) \\
& - d\cos\phi_0\cos\theta_0(\epsilon^n + \epsilon^{n+1})\frac{L}{2}\left(\frac{\Phi_b''(\mathbf{r}_{ba}^{n+})}{|\mathbf{r}_{ba}^{n+}|^2} - \frac{\Phi_b'(\mathbf{r}_{ba}^{n+})}{|\mathbf{r}_{ba}^{n+}|^3}\right)(r_{ba,x}^{n+}\cos\phi_0\cos\theta_0 + r_{ba,y}^{n+}\sin\phi_0\cos\theta_0 - r_{ba,z}^{n+}\sin\theta_0) \\
& + d\cos\phi_0\cos\theta_0(\epsilon^n + \epsilon^{n-1})\frac{L}{2}\left(\frac{\Phi_b''(\mathbf{r}_{ba}^{n-})}{|\mathbf{r}_{ba}^{n-}|^2} - \frac{\Phi_b'(\mathbf{r}_{ba}^{n-})}{|\mathbf{r}_{ba}^{n-}|^3}\right)(r_{ba,x}^{n-}\cos\phi_0\cos\theta_0 + r_{ba,y}^{n-}\sin\phi_0\cos\theta_0 - r_{ba,z}^{n-}\sin\theta_0) \\
& - d\cos\phi_0\cos\theta_0(\eta^n + \eta^{n+1})\frac{L}{2}\left(\frac{\Phi_b''(\mathbf{r}_{ba}^{n+})}{|\mathbf{r}_{ba}^{n+}|^2} - \frac{\Phi_b'(\mathbf{r}_{ba}^{n+})}{|\mathbf{r}_{ba}^{n+}|^3}\right)(-r_{ba,x}^{n+}\sin\phi_0\sin\theta_0 + r_{ba,y}^{n+}\cos\phi_0\sin\theta_0) \\
& + d\cos\phi_0\cos\theta_0(\eta^n + \eta^{n-1})\frac{L}{2}\left(\frac{\Phi_b''(\mathbf{r}_{ba}^{n-})}{|\mathbf{r}_{ba}^{n-}|^2} - \frac{\Phi_b'(\mathbf{r}_{ba}^{n-})}{|\mathbf{r}_{ba}^{n-}|^3}\right)(-r_{ba,x}^{n-}\sin\phi_0\sin\theta_0 + r_{ba,y}^{n-}\cos\phi_0\sin\theta_0)\Bigg] \\
& + \frac{Q_b L}{2}(a \leftrightarrow b) - Q_a\frac{\partial \Phi_{\text{ext}}(\mathbf{a}^n)}{\partial \theta^n} - Q_b\frac{\partial \Phi_{\text{ext}}(\mathbf{b}^n)}{\partial \theta^n}
\end{aligned}$$

where each of the above terms, such as $\mathbf{r}_{ba}^{n+}$, is to be evaluated at $(\theta_0, \phi_0)$.